\documentclass[a4paper,twoside]{article}

\newcommand{\eg}{e.g.,\xspace}
\newcommand{\etc}{etc.\xspace}

\newcommand{\ie}{i.e.,\xspace}
\usepackage{xspace}
\usepackage{framed}
\usepackage{booktabs}
\usepackage{listings}
\usepackage{xcolor}

\usepackage{url}
\usepackage{epsfig}
\usepackage{subcaption}
\usepackage{calc}
\usepackage{amssymb}
\usepackage{amstext}
\usepackage{amsmath}
\usepackage{amsthm}
\usepackage{multicol}
\usepackage{pslatex}
\usepackage{apalike}
\usepackage{algorithm2e}
\usepackage[bottom]{footmisc}

\usepackage{SCITEPRESS}

\begin{document}

\title{I Know What You Bought Last Summer: Investigating User Data Leakage in E-Commerce Platforms}

\author{\authorname{Ioannis Vlachogiannakis\sup{1,2}, Emmanouil Papadogiannakis\sup{1,2}, Panagiotis Papadopoulos\sup{1}, \\ Nicolas Kourtellis\sup{3} and Evangelos Markatos\sup{1,2}}
\affiliation{\sup{1}Foundation for Research and Technology - Hellas (FORTH), Heraklion, Greece}
\affiliation{\sup{2}University of Crete, Heraklion, Greece}
\affiliation{\sup{3}Telefonica Research, Barcelona, Spain}
}

\keywords{Private Information Leakage, Sensitive User Data, E-Commerce, Privacy, Tracking}
\abstract{
In the digital age, e-commerce has transformed the way consumers shop, offering convenience and accessibility.
Nevertheless, concerns about the privacy and security of personal information shared on these platforms have risen. 
In this work, we investigate user privacy violations, noting the risks of data leakage to third-party entities.
Utilizing a semi-automated data collection approach, we examine a selection of popular online e-shops, revealing that nearly 30\% of them violate user privacy by disclosing personal information to third parties.
We unveil how minimal user interaction across multiple e-commerce websites can result in a comprehensive privacy breach.
We observe significant data-sharing patterns with platforms like Facebook, which use personal information to build user profiles and link them to social media accounts.
}

\onecolumn \maketitle \normalsize \setcounter{footnote}{0} \vfill

\section{\uppercase{Introduction}}
\label{sec:introduction}

According to studies~\cite{forbesOnlineShopping}, 34\% of shoppers shop online at least once a week.
In general, the e-commerce market is expected to reach a staggering \$8 trillion value by 2027.
Apart from the apparent comfort of shopping without visiting physical stores, one of the key factors of the rapid growth of e-commerce has been the extensive use of data and third-parties (\ie analytics, media buttons, advertising networks).

By integrating such third-party services, e-shops can optimize inventory levels by analyzing historical sales data, identifying seasonal trends, and forecasting future demand.
Additionally, and more importantly, by collecting and processing a great wealth of user and behavioral data, third-party analytics can provide a unique understanding of customer behavior and their purchase patterns~\cite{dpgmedia}.
This allows e-shops to increase customer retention and perform targeted advertising~\cite{visaTargeted}.

In some cases, e-shops, may not have full control or even awareness of what/how pervasive the tracking of the third-party tools they embed in their platforms is.
On the other hand, customers can only trust their sensitive personal information (\eg contact details and payment information) to e-shops and expect that this information will be used only by them and for the sole purpose of purchasing products or services.

Unfortunately, there are various examples where this trust was breached.
In 2025, an Austrian privacy non-profit filed complaints accusing e-shops like AliExpress, SHEIN, Temu for violating data protection regulations in the European Union by unlawfully transferring users' data to China~\cite{tiktokAliexpress}.
In 2024 the state of Arkansas sued the Chinese online retailer Temu for illegally accessing user information~\cite{temuSued}.
In 2023, hundreds of online stores were reported for accidentally leaking customer data in public folders without any restrictions~\cite{backupLeaks}. 
In 2019, a study showed that at least 80\% of shopping apps leak users' data~\cite{shoppingApps}.

Prior research on e-commerce has highlighted several vulnerabilities in online shopping platforms, including weak API security~\cite{floresexamining}, third-party tracking~\cite{rauti2024analyzing}, and insufficient data protection measures~\cite{pagey2023all}.
Previous works have identified that e-commerce platforms share user data, but do not provide a comprehensive mapping of how information flows between different third-party entities.
Data leaks are often examined in isolation (\ie individual e-shop platforms), 
ignoring the aggregation of user data across multiple platforms. 
A more comprehensive
approach is essential to map the full lifecycle of leaked data globally.

In this work, we investigate the leakage of sensitive user information from e-commerce platforms to third parties and explore how these entities can aggregate user data across multiple e-shops.
We explore whether the personal information that users provide to e-commerce platforms are shared with third parties, contrarily to the user's expectations.
Our findings highlight that privacy leaks are not limited to obscure sites, but extend to highly popular e-shops with million of monthly visitors.
Additionally, even limited interaction with multiple platforms can lead to complete exposure of a user's personal information.

The main contributions of this work are:
\begin{enumerate}
    \item We compare the data-leaking behaviors of e-shops with those in other industries, emphasizing the heightened privacy risks in the e-commerce and shopping field.
    \item We discover that 29\% of the online retail stores in our dataset share at least one piece of their users' sensitive private information to a third-party entity.
    We highlight that this behavior is evident even in extremely popular platforms with millions of monthly visitors.
    \item We demonstrate that third parties are capable of aggregating personal information from multiple e-commerce platforms to construct comprehensive user profiles.
    In fact, users engaging with as few as five e-commerce sites may have their entire profile exposed to third parties.
    \item We reveal that Meta is the third party that receives the largest amount of private information, allowing it to use this information to match shopping behaviors with Facebook accounts.
    \item We make our tools and dataset publicly available to foster further research on the field.
\end{enumerate}
\begin{figure*}
    \centering
    \includegraphics[width=\textwidth]{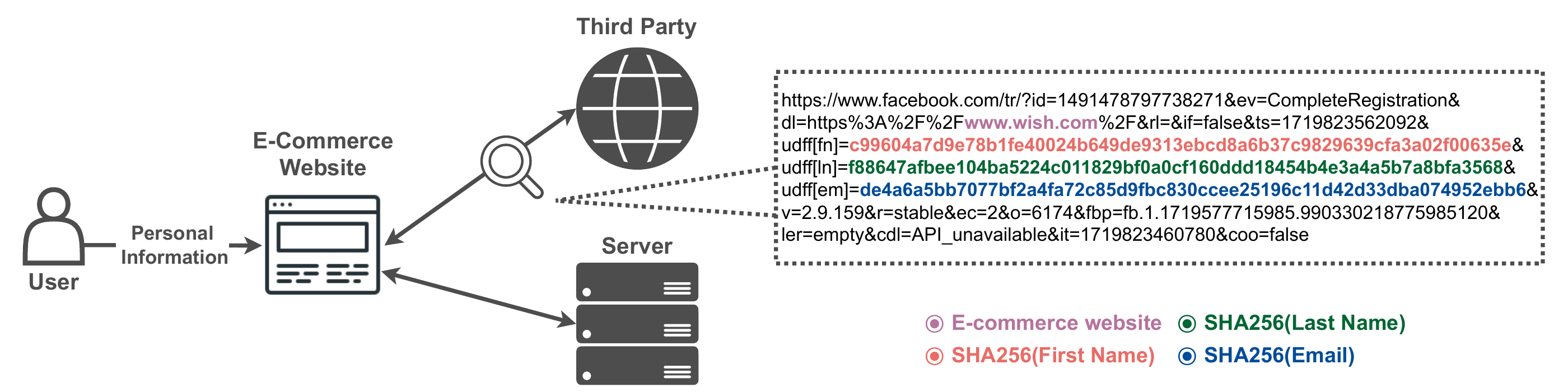}
    \caption{Overview of methodology for detecting personal information leakage.}
    \label{fig:methodology}
\end{figure*}

\section{\uppercase{Methodology}}
\label{sec:methodology}

In this work, we follow a two-phase methodology.
The first phase involves data collection, where we emulate real-world scenarios to gather data from e-commerce platforms.
The second phase involves data analysis, where we process the collected information.
For data collection, we develop a semi-automated crawler using the Playwright framework~\cite{playwright}, to systematically extract data from the websites of selected e-shop platforms.

\subsection{Data Collection}
\label{sec:dataCollection}

First, we compile a list of popular and representative e-commerce websites from around the world.
To that extent, we utilize SimilarWeb~\cite{similarweb} to gather the most popular five e-shop platforms in each country along with the top 50 worldwide, and accumulate a list of 200 distinct e-commerce websites.
We make our list publicly available to foster further research on the field~\cite{openSource}.
Then, we visit all the e-commerce platforms with our semi-automated crawler, 
located in an EU institute.

We build our tool to manage a browser instance in a way that it collects all network traffic and cookie jar (both first-party and third-party cookies), as the user interacts with the browser.
Specifically, for e-shop websites, these actions involve creating a user account, browsing products, and preparing for a purchase.
Our scenario simulates real user activities such as product search, adding items to the cart, and progressing through steps leading up to payment and order confirmation.
We deliberately refrain from completing any purchases to avoid impacting the platforms or merchants in any way.
Ethical considerations are further discussed in Section~\ref{subsec:ethical_considerations}.

Our goal is to emulate a real-world case reflecting a typical user with unique personal information such as full name, mobile phone, email address, physical address, etc.
Towards that extent, we create a fake persona of a user from the country where the crawler is located.
It is important to highlight that the persona is consistent and in each platform we provide the same personal information in the same manner as a real user.
Our persona consists of the following personal and sensitive information:
(1) email,
(2) name, 
(3) phone number, 
(4) gender, 
(5) zip code,
(6) credit card details,
(7) username,
and (8) password.

\subsection{Leakage Detection}

We manually visit all the websites in our list during July 2024 independently of each other, starting with a clean browser context for each e-commerce platform.
We extensively study the network traffic generated during our visit to each e-commerce platform.
To detect e-commerce websites that leak sensitive information, we study how information flows from websites to third parties.
We group HTTP(S) requests and filter them by destination URLs to identify all third parties, with which the platforms communicate.
Additionally, we make use of the DuckDuckGo Tracker Radar dataset~\cite{tracker_radar} to match each one of the third-party domains with the company that owns or operates them, aiming to link information leaks with specific companies or larger legal entities.

Finally, we iterate through all HTTP(S) requests and third-party cookies, searching for occurrences of all sensitive personal information that we inserted when creating each profile.
We search for this information in different formats, including plaintext, SHA256 or MD5 Hashed, URL Encoded and Base64 Encoded.
We provide an overview of our methodology in Figure~\ref{fig:methodology}, where we demonstrate how we capture personal information being shared with third parties.
We conduct an in-depth analysis of the collected data to unveil matches in the URL parameters of GET requests and the body of POST requests, as well as in each cookie value. 
We study information flows between the client browser and third parties, acknowledging that data sent directly from the store's server to third parties is not captured by our methodology.

Website operators might claim that sharing hashed personal data with third parties (as shown in Figure~\ref{fig:methodology}) is harmless, since the original data cannot be extracted.
However, we argue that this is a misleading argument.
Third parties, such as analytics services and social networks, with access to vast amounts of user data, can easily link hashed values to their own databases.
If a service already holds a user's email address, it can effortlessly identify the individual by matching the hash to its database entry.
\section{\uppercase{Sensitive Information Leakage}}
\label{sec:findings}

Inspired by anecdotal evidence that e-commerce platforms collect an extensive amount of user information~\cite{temuSued,dpgmedia,visaTargeted}, we perform a preliminary investigation to understand to what extent this is happening.
Our goal is to discover indications that private information leaks are more common in e-commerce platforms since they have access to more user information compared to other types of websites.
We develop a data collection tool designed to automatically visit websites and capture network traffic along with the cookie jar, including both first-party and third-party cookies.
The tool navigates to a website's landing page and waits for the page to fully load before collecting the relevant data.
We select five different categories of websites to study,
(i) ``E-commerce and Shopping'', 
(ii) ``Business and Consumer services'',
(iii) ``Health'',
(iv) ``Travel and Tourism'',
and (v) ``Finance''.
For this experiment, we process 200 websites from each category extracted from SimilarWeb~\cite{similarweb}.
We make the lists of websites per category publicly available~\cite{openSource}.
We collect all requests and cookies, extract third-party entities, and compare the number of interactions to determine which category engages with the highest number of third parties.

\begin{table}[h!]
    \caption{Average, Median and 90th Percentile of third-party interactions. E-commerce websites interact with more third-party services than other categories.}
    \resizebox{\columnwidth}{!}{
    \begin{tabular}{lrrr}
    \toprule
        \textbf{Category of Websites} & \textbf{Average} & \textbf{Median} & \textbf{90\textsuperscript{th} Percentile}\\
    \midrule
        E-commerce \& Shopping & 17 & 12 & 40\\
        Business \& Consumer & 15 & 12 & 33\\
        Health & 15 & 12 & 34\\
        Travel \& Tourism & 14 & 10 & 35\\
        Finance & 13 & 11 & 31\\
    \bottomrule
    \end{tabular}
    }
    \label{tab:average_median_numbers}
\end{table}

Our analysis reveals that websites in the ``E-commerce and Shopping'' category interact with more third-party services than those in other categories.
We present our findings in Table~\ref{tab:average_median_numbers}, showing the average, median and 90\textsuperscript{th} percentile of third-party interactions across five categories.
A deeper analysis of e-commerce platforms from our original list (see Section~\ref{sec:dataCollection}) reveals that the average and median number of third parties can rise to 21 and 14, respectively, when users spend more time on the website, interacting with its components and navigating to more pages apart from the landing page.
Previous work has already demonstrated that landing and internal pages can have significant differences in the number of trackers~\cite{10.1145/3419394.3423626}.

\begin{leftbar}
\noindent\textbf{Finding 1:}
E-commerce websites engage with more third-party services than other categories, with the number of interactions increasing as users spend more time navigating the site.
\end{leftbar}

\begin{figure}[ht]
    \centering
    \includegraphics[width=1\columnwidth]{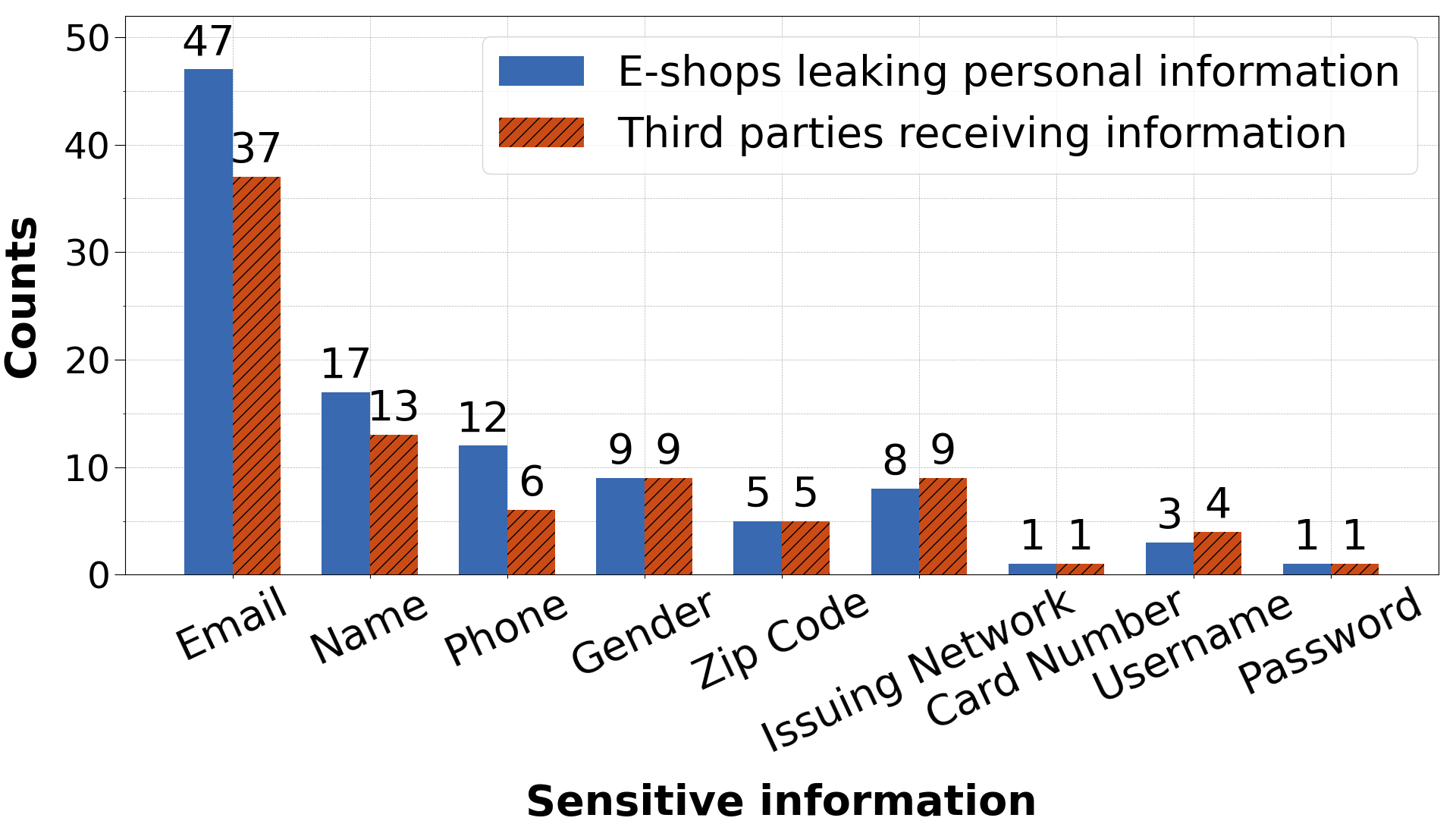}
    \caption{Number of e-shops leaking sensitive personal information and number of third parties collecting this information from different e-commerce platforms.}
    \label{fig:leaking_receiving}
\end{figure}

\subsection{Sensitive Data Flows}
\label{subsec:general_results}

We analyze the collected data following the methodology outlined in Section~\ref{sec:methodology} to identify instances of sensitive information leakage from e-commerce platforms to third parties.
We discover that 57 out of 200 digital shopping platforms we investigated, almost 30\%, leak at least one piece of sensitive user information to an external legal entity.
This means that almost one in three online stores transmits sensitive user data, either encoded or in plain text, to unrelated third parties.
Unlike pseudonymous tracking methods such as third-party cookies or browser fingerprinting~\cite{10.1145/3442381.3450056}, this type of data leakage is particularly concerning because it involves personally identifiable information (PII), including full names, email addresses, and physical addresses.
The collection of such information from third parties not only compromises user privacy, by enabling detailed profiling, but also increases the risk of exposure in cases of data breaches, a common occurrence in the last few years (\eg~\cite{npdBreach,amazonBreach}).

In Figure~\ref{fig:leaking_receiving}, we illustrate the number of e-commerce platforms sharing personal information (blue bars), as well as the third party entities collecting user information from various e-commerce platforms (red bars).
\begin{figure}[ht]
    \centering
    \includegraphics[width=1\columnwidth]{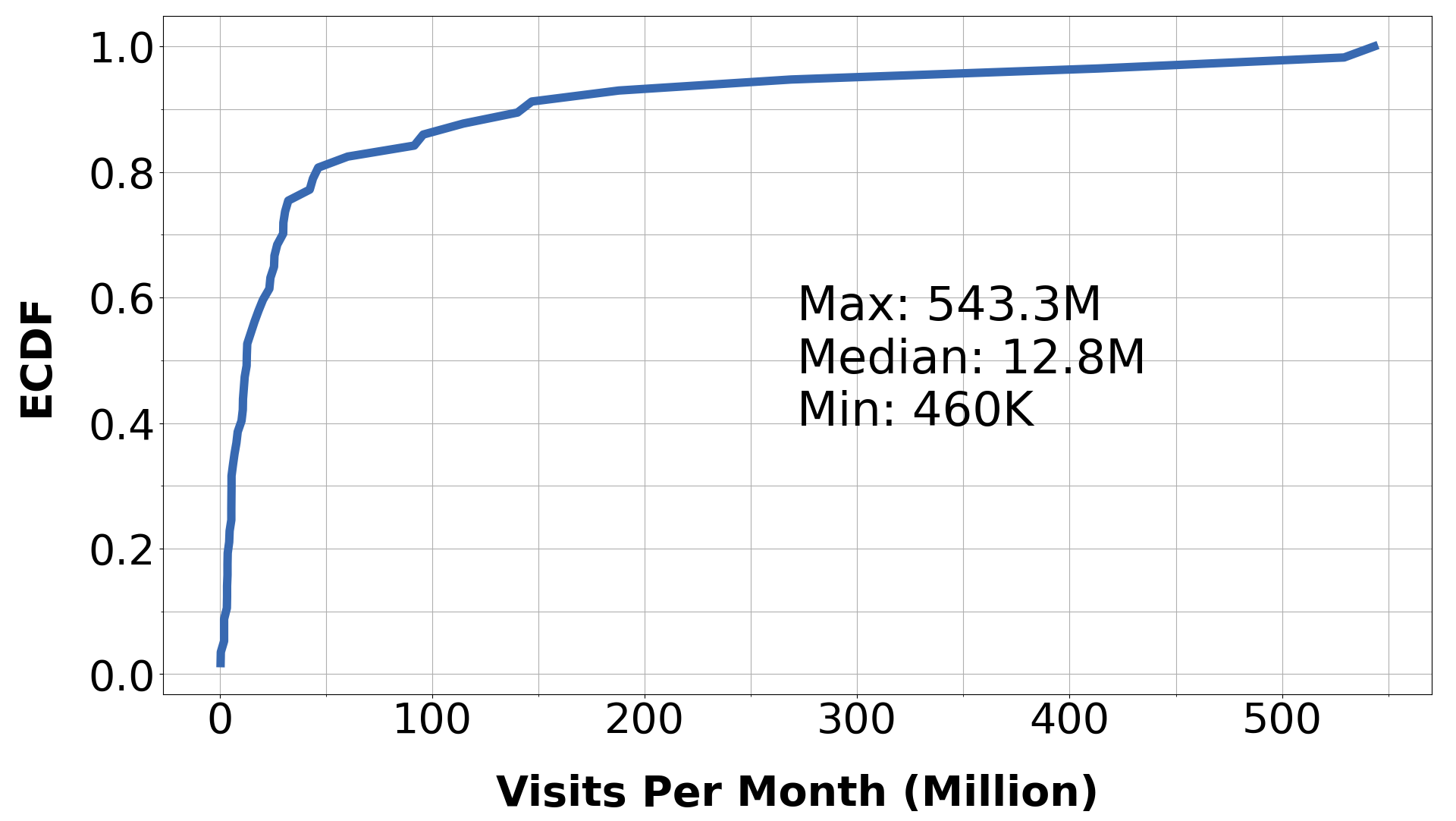}
    \caption{Distribution of monthly visits (both desktop and mobile) of e-commerce websites.}
    \label{fig:popularity_distribution}
\end{figure}
We observe that the email address, a piece of sensitive information that uniquely identifies a user, is leaked from 47 online stores.
Also, it is worth noting that 37 distinct third parties collect this information and can, at the very least, correlate where and when a specific user (\ie email address) shops online.
These third parties include popular conglomerates that provide analytics services (\eg Facebook, Google) as well as companies that can correlate shopping profiles with user accounts in other platforms (\eg ByteDance Ltd that developed TikTok).
Third-party data brokers buy and compile information from multiple sources, often without users’ knowledge.
If separate e-commerce platforms leak partial user information, data brokers can combine these fragments to create complete profiles, including names, addresses, purchase history, and even preferences.

Next, we study the popularity of the 57 retail online platforms leaking user information. 
Figure~\ref{fig:popularity_distribution} illustrates their popularity distribution, which is based on number of monthly visits, both from mobile and desktop clients, obtained from SimilarWeb.
Our analysis reveals that e-commerce platforms leaking user information range from low-visibility websites to those with substantial monthly traffic.
While it is somewhat expected that less popular online stores may engage in such practices due to limited regulatory oversight, we also observe this behavior in highly popular platforms with millions of monthly visitors, including AliExpress and Etsy.
Altogether, e-commerce platforms that share at least one piece of personal information have an aggregated traffic of \emph{3.23} Billion monthly visits, significantly increasing the risk of user data exposure on a massive scale.

\begin{leftbar}
\noindent\textbf{Finding 2:}
Our analysis reveals that 29\% of the online retail stores in our dataset, including highly popular platforms with millions of monthly visitors, leak at least one piece of their users’ sensitive private information.
\end{leftbar}

\subsection{Data Aggregation}

\begin{figure}[ht]
    \centering
    \includegraphics[width=1\columnwidth]{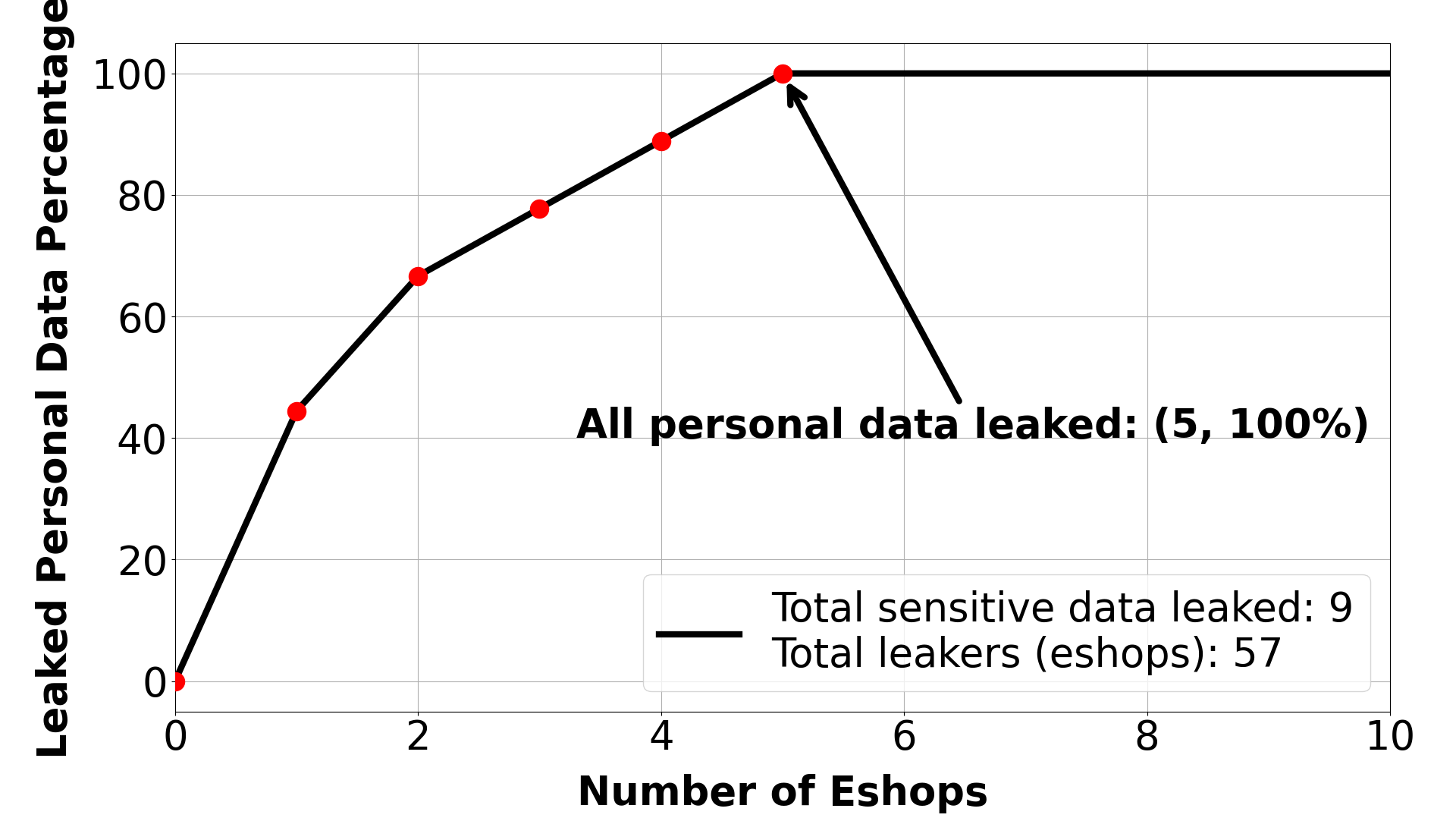}
    \caption{Complete exposure of a user's personal information when visiting as few as 5 e-shop platforms.}
    \label{fig:exposure_to_trackers}
\end{figure}

Numerous third-party tracking services, such as Google Analytics, Meta Pixel, and various advertising networks, aggregate data from multiple e-commerce platforms. When a user engages with sites like AliExpress and Wayfair, these tracking entities can correlate their activity across platforms, enabling the construction of comprehensive consumer profiles.
We discover that a user visiting as few as five different e-shop platforms, can have their entire personal profile (including contact information, account credentials, payment details, \etc) shared with third parties, as illustrated in Figure~\ref{fig:exposure_to_trackers}.
This personal information, which includes sensitive fields such as email address, name, and phone number collectively forms a comprehensive user profile, as described in Section~\ref{sec:dataCollection}.
This suggests that even limited interactions with e-shops can pose significant privacy risks, as users are unknowingly subjected to data sharing without their explicit consent.

\begin{figure*}[ht]
    \centering
    \includegraphics[width=0.78\textwidth]{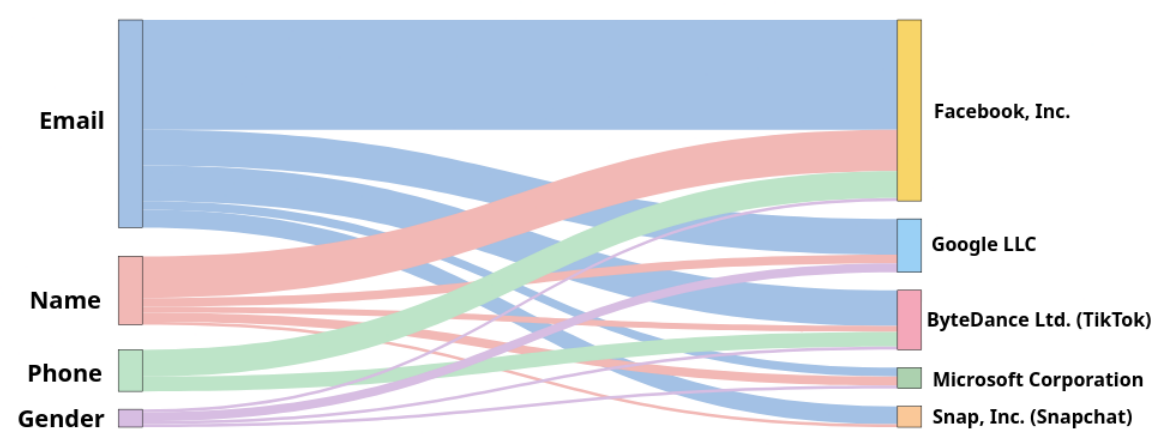}
    \caption{Information flow of sensitive personal information that e-commerce platforms distribute to third-party entities. A greater flow weight indicates that a third party receives information from multiple online stores.}
    \label{fig:sankey_diagram}
\end{figure*}

We aggregate popular third-party entities that receive personal information from the e-shops we study, and present them in Table~\ref{tab:companies_array}, along with the information they collect.
It is evident that Web conglomerates such as \emph{Facebook}, \emph{Google} and \emph{Microsoft} not only collect very sensitive personal information, but they do so from multiple online stores, thus tracking users when they shop in various platforms.
We illustrate the flow of sensitive user information towards third parties in Figure~\ref{fig:sankey_diagram}.
This plot illustrates the most critical fields of personal information and the third parties that receive this data.
Each flow, represented by its width, indicates the volume of e-shops leaking a piece of personal information to a third party.
A wider flow suggests a higher number of e-commerce platforms sharing sensitive data.
We observe that the email address, a unique identifier, is commonly leaked by e-shops.
Moreover, it is evident that Facebook collects the most data from e-commerce platforms.

\begin{table}[!ht]
    \caption{Third-party legal entities acquiring the most personal information from multiple e-commerce platforms. Each cell represents the number of distinct e-shops sharing specific personal information with each third-party entity.}
    \centering
    \resizebox{\columnwidth}{!}{
    \begin{tabular}{lrrrr}
    \toprule
        \textbf{Third-Party Company} & \textbf{Email} & \textbf{Name} & \textbf{Phone} & \textbf{Gender} \\
    \midrule
        Facebook, Inc. & 37 & 14 & 9 & 1 \\
        Google LLC & 12 & 3 & - & 3 \\
        ByteDance Ltd. (TikTok) & 12 & 2 & 5 & 1 \\
        Microsoft Corporation & 3 & 3 & - & 1 \\
        Snap Inc. (Snapchat) & 6 & 1 & - & - \\
    \bottomrule
    \end{tabular}
    }
    \label{tab:companies_array}
\end{table}

It is worth noting that when emulating the real world scenario described in Section~\ref{sec:methodology}, we browse the same category of products wherever possible (\eg shoes) in each platform and add them to the virtual cart.
We discover that some e-shop platforms inform third parties about the products a user is interested in.
In Listing~\ref{lst:facebook_pr_leak_url}, we demonstrate a case of a decoded URL informing Facebook that the user browsed for a specific category of products.

Many e-commerce platforms share user data with third-party trackers, which operate across multiple websites.
These trackers can link user activity from different platforms to create detailed consumer profiles, including sensitive information like names, addresses, and purchase history - often without users' awareness.
This creates significant privacy risks, allowing third parties to build invasive profiles of their online activities and personal preferences

\begin{leftbar}
\noindent\textbf{Finding 3:}
Users interacting with as few as five e-commerce sites risk having their entire profile exposed to third parties that consolidate personal information to create detailed user profiles.
\end{leftbar}

\lstset{
    basicstyle=\ttfamily\small,     
    breaklines=true,    
    frame=single,                
    keywordstyle=\color{red},          
    escapeinside={(*@}{@*)},
    linewidth=\textwidth,
    captionpos=b,
    xleftmargin=10pt,
    xrightmargin=10pt,
    framexleftmargin=10pt,
    framexrightmargin=10pt
}

\begin{figure*}[t]
    \normalsize
    \centering
    \begin{lstlisting}[caption={Destination URL captured in one of the requests in our dataset. Its destination is the tracking service of Facebook and as a parameter is passed the exact url of the product that our virtual user visited.}, label={lst:facebook_pr_leak_url}, breaklines=true,xleftmargin=0.5cm,xrightmargin=0.5cm]
https://www.facebook.com/tr/?id=693212574061933&ev=PageView&dl=https://www.thewarehouse.co.nz/c/clothing-shoes-accessories/(*@\textcolor{red}{mens-clothing-shoes}@*)/(*@\textcolor{red}{mens-shoes}@*)/(*@\textcolor{red}{mens-sports-hiking-shoes}@*)&rl=https://www.thewarehouse.co.nz/account&if=false&ts=1719495540857&sw=1920&sh=1200&ud[em]=de4a6a5bb7077bf2a4fa72c
85d9fbc830ccee25196c11d42d33dba074952ebb6&v=2.9.159&r=stable&a=tmSimo-GTM-WebTemplate&ec=0&o=4126&fbp=fb.2.1719495152216.77178234915891118&ler=empty&it=1719495540626&coo=false&eid=1719495539212_1719496172506_1_gtm.js&tm=1
    \end{lstlisting}
\end{figure*}

\section{Effective Persistent Tracking}
\label{subsec:facebook}

We observe in Figure~\ref{fig:sankey_diagram} that there is a significant flow of sensitive information from retail platforms towards Facebook.
Upon closer inspection, we find that 18.5\% of the e-shop platforms we studied leak the user's email address to Facebook.
This information is shared to Facebook as a hashed value, often along other fields like username.
Facebook, one of the largest social networks, has an extensive collection of user email addresses.
Through tracking on e-shops, Facebook can link users browsing e-shops to specific Facebook accounts.
This form of tracking is particularly effective, as email addresses (unlike  pseudonymous 
third-party cookies) are unique and directly identifiable.

To make matters worse, we observe that when e-shop platforms send requests towards Facebook's endpoints, the product or category of products that the user is browsing is also leaked (Listing~\ref{lst:facebook_pr_leak_url}).
Facebook is, therefore, capable of tracking users' shopping behavior by identifying products they have seen or bought and build an extensive user profiles.
As a result, Facebook not only \emph{knows where and when you are shopping, but also what you are shopping for}.

We argue that this form of tracking has a less apparent dimension.
Facebook is able to track users through its tracking services, that retail online stores integrate to their platforms, even without owning a Facebook account.
Once the user decides to create an account, Facebook can associate all previously collected personal and historical data with the newly created profile.
This enables Facebook to gain insights into the \emph{user's past shopping habits and preferences}.
As a result, the company acquires a comprehensive understanding of consumer behavior, allowing it to offer personalized recommendations and advertisements.
To put it into perspective, when a new user creates an entirely new account with Facebook, the company may already be aware of their shopping habits.
This behavior has been similarly noted in previous research involving Facebook's Pixel tracking technology~\cite{bekos2023hitchhiker}.

Finally, Facebook is part of the Meta group, which also operates Instagram and WhatsApp, thus broadening the scope of data collection across multiple social platforms.
This interconnected network grants Meta comprehensive insights into user preferences and shopping behaviors.
By leveraging data from Facebook, WhatsApp, and Instagram, Meta can effectively track a diverse range of users, often segmented by age groups linked to each platform~\cite{socialMediaFacts}.
The cumulative traffic across the 37 e-shops, that disclose users' emails to Facebook, is \emph{2.35} Billion, highlighting the substantial reach and potential privacy impact of these data sharing practices.

\begin{leftbar}
\noindent\textbf{Finding 4:}
Meta is the third party receiving the most significant amount of private information, enabling the company to correlate shopping behaviors with specific Facebook accounts.
\end{leftbar}
\section{\uppercase{Related Work}}
\label{sec:related_work}

In~\cite{okeke2013issues}, the authors highlighted privacy and trust concerns among online customers regarding data security and sharing personal information with third parties.
In~\cite{gurung2016online}, the authors suggest that privacy concerns have a greater impact on risk assessment than security concerns, influencing consumer attitudes and intentions toward online shopping.
In~\cite{broeder2020culture}, the authors found that privacy notices indirectly influenced trust and purchase behavior by assuring consumers of personal information protection.
Gaining their satisfaction and trust leads customers to prioritize online shopping against traditional shopping methods, as authors in~\cite{kurniawan2024importance} discuss.
In~\cite{martiskova2020digital} researchers
revealed that both genders are equally willing to deny a purchase, due to extreme personal data requirements.
In~\cite{10.1145/3359183}, the authors investigated the prevalence of deceptive design practices in 11K popular shopping websites, discovering that about 11.1\% displayed at least one instance of dark patterns.

In~\cite{pabian2020customer}, authors identified key security threats related to payment methods, personal data, and purchased goods for both customers and sellers.
Researchers in~\cite{degutis2023consumers} indicated
that consumers value the expected give-and-take from e-commerce providers more than the direct benefits of data disclosure.
Authors in~\cite{diaz2016privacy} demonstrated that privacy threats are present in all stages of the e-shopping process, thus protecting only individual stages is insufficient.
On top of that, in~\cite{di2020toward}, the authors discussed the challenges of balancing data availability for analysis with individual control over personal data. 
In~\cite{moric2024protection}, researchers emphasize the importance of robust data security measures in e-shops, presenting a framework that integrates legal, technological, and procedural elements to enhance data protection and consumer trust, aligned with standards like GDPR.

In addition, the impact of Secure Multiparty Computing (MPC) on traditional factors such as control, trust, and risk in data sharing decisions enhances control, reduces the need for interorganizational trust and prevents data leakage~\cite{agahari2022not}.
At the same time, MPC enables a ``privacy-as-a-service'' business model, enhancing security and reducing trust dependencies on data marketers, while providing new revenue opportunities through analytics and privacy services~\cite{agahari2021business}.
Researchers in~\cite{sakalauskas2024personalized} introduced an algorithm, which uses clickstream data for targeted advertising to high-value customers.
By measuring user activity, advertisers will improve ad performance,
while costs can be reduced.
\section{\uppercase{Conclusion \& Discussion}}
\label{sec:conclusion}

\subsection{Summary}
\label{subsec:summary}

In this work, we explore user privacy breaches on e-shop platforms in a global scale.
We find that e-commerce websites interact with the most third-party entities, suggesting that there is a potential leak of private information towards third parties.
In fact, we study 200 distinct e-shops platforms from countries around the world and discover that nearly 30\% of these leak at least one piece of sensitive information to a third-party entity.
In addition, we find that Web conglomerates such as Facebook collect sensitive user information from multiple e-shops, and that they can use this information to match shopping habits with online user profiles.
Finally, we highlight that even minimal interactions with these platforms can lead to substantial privacy risks, as a profile can be compromised after engaging with just five online stores.
These findings highlight the need to take protective measures, enhance privacy protection and transparency in handling data over retail online shops. 

\subsection{Discussion}
\label{subsec:discussion}

The findings of this work emphasize the need for improved transparency, privacy, and trust regarding personal data.
While platforms likely disclose the sharing of sensitive information in their lengthy and obscure terms and conditions, there is a mismatch with user expectations when registering on online stores.
Users typically assume their sensitive information will only be used for purchasing products, not shared with unknown third parties.
Consumers are increasingly aware of privacy risks, which may influence their trust and shopping behavior in online shops.

\subsection{Ethical Considerations}
\label{subsec:ethical_considerations}

In this study, we made deliberate efforts to study the e-commerce ecosystem without disrupting it.
The data collection process consisted of manual actions, minimizing the use of instrumented operations to a minimal.
In Section~\ref{sec:findings}, our automated system visited only the landing page of each website to assess third-party interactions, ensuring no impact on its performance.
Each website was processed once, one at a time, simulating real user activity.
Lastly, no personal data was collected or shared,
adhering to research ethics principles~\cite{rivers2014ethicalresearchstandards}. 

\bibliographystyle{apalike}
{\small
\def\UrlBreaks{\do\/\do-\do_}
\bibliography{main}}

\begin{thebibliography}{}

\bibitem[Agahari et~al., 2021]{agahari2021business}
Agahari, W., Dolci, R., and de~Reuver, G. (2021).
\newblock Business model implications of privacy-preserving technologies in data marketplaces: The case of multi-party computation.
\newblock In {\em 29th European Conference on Information Systems (ECIS 2021) A Virtual AIS Conference: Human Values Crisis in a Digitizing World}, pages 1--16. Association of the Information Systems.

\bibitem[Agahari et~al., 2022]{agahari2022not}
Agahari, W., Ofe, H., and de~Reuver, M. (2022).
\newblock It is not (only) about privacy: How multi-party computation redefines control, trust, and risk in data sharing.
\newblock {\em Electronic markets}, 32(3):1577--1602.

\bibitem[Aqeel et~al., 2020]{10.1145/3419394.3423626}
Aqeel, W., Chandrasekaran, B., Feldmann, A., and Maggs, B.~M. (2020).
\newblock On landing and internal web pages: The strange case of jekyll and hyde in web performance measurement.
\newblock In {\em Proceedings of the ACM Internet Measurement Conference}, page 680–695, New York, NY, USA. Association for Computing Machinery.

\bibitem[Bekos et~al., 2023]{bekos2023hitchhiker}
Bekos, P., Papadopoulos, P., Markatos, E.~P., and Kourtellis, N. (2023).
\newblock The hitchhiker’s guide to facebook web tracking with invisible pixels and click ids.
\newblock In {\em Proceedings of the ACM Web Conference 2023}, pages 2132--2143.

\bibitem[Broeder, 2020]{broeder2020culture}
Broeder, P. (2020).
\newblock Culture, privacy, and trust in e-commerce.
\newblock {\em Marketing from Information to Decision Journal}, 3(1):14--26.

\bibitem[Center, 2024]{socialMediaFacts}
Center, P.~R. (2024).
\newblock Social media and news fact sheet.
\newblock \url{https://www.pewresearch.org/journalism/fact-sheet/social-media-and-news-fact-sheet/}.

\bibitem[Degutis et~al., 2023]{degutis2023consumers}
Degutis, M., Urbonavi{\v{c}}ius, S., Hollebeek, L.~D., and Anselmsson, J. (2023).
\newblock Consumers’ willingness to disclose their personal data in e-commerce: A reciprocity-based social exchange perspective.
\newblock {\em Journal of Retailing and Consumer Services}, 74:103385.

\bibitem[di~Vimercati et~al., 2020]{di2020toward}
di~Vimercati, S. D.~C., Foresti, S., Livraga, G., and Samarati, P. (2020).
\newblock Toward owners’ control in digital data markets.
\newblock {\em IEEE Systems Journal}, 15(1):1299--1306.

\bibitem[Diaz et~al., 2016]{diaz2016privacy}
Diaz, J., Choi, S.~G., Arroyo, D., Keromytis, A.~D., Rodriguez, F.~B., and Yung, M. (2016).
\newblock Privacy threats in e-shopping.
\newblock In {\em Data Privacy Management, and Security Assurance: 10th International Workshop, DPM 2015, and 4th International Workshop, QASA 2015, Vienna, Austria, September 21--22, 2015. Revised Selected Papers 10}, pages 217--225. Springer.

\bibitem[DuckDuckGo, 2020]{tracker_radar}
DuckDuckGo (2020).
\newblock Duckduckgo tracker radar.
\newblock \url{https://github.com/duckduckgo/tracker-radar}.

\bibitem[Fabbro, 2024]{visaTargeted}
Fabbro, R. (2024).
\newblock Visa will give customer data to retailers for ai-targeted ads.
\newblock \url{https://qz.com/visa-data-tokens-ai-marketing-retailers-transactions-1851481158}.

\bibitem[Firstpost, 2024]{amazonBreach}
Firstpost (2024).
\newblock Amazon confirms security breach where employee data of millions was leaked.
\newblock \url{https://www.firstpost.com/tech/amazon-confirms-security-breach-where-employee-data-of-millions-was-leaked-13834672.html}.

\bibitem[Flores et~al., 2022]{floresexamining}
Flores, R., Perine, C., Remorin, L., and Reyes, R. (2022).
\newblock Examining security risks in logistics apis used by online shopping platforms.
\newblock \url{https://www.trendmicro.com/vinfo/us/security/news/online-privacy/pii-leaks-and-other-risks-from-unsecure-e-commerce-apis}.

\bibitem[Forbes, 2024]{forbesOnlineShopping}
Forbes (2024).
\newblock 35 e-commerce statistics of 2024.
\newblock \url{https://www.forbes.com/advisor/business/ecommerce-statistics/}.

\bibitem[Group, 2024]{dpgmedia}
Group, D.~M. (2024).
\newblock Dpg media belgium extends digital advertising offering with exclusive retail data from carrefour via unlimitail and tom\&co.
\newblock \url{https://www.dpgmediagroup.com/extension-digital-advertising-offering-retaildata-carrefour-unlimitail-tomenco}.

\bibitem[Gurung and Raja, 2016]{gurung2016online}
Gurung, A. and Raja, M.~K. (2016).
\newblock Online privacy and security concerns of consumers.
\newblock {\em Information \& Computer Security}, 24(4):348--371.

\bibitem[Klappholz, 2024]{npdBreach}
Klappholz, S. (2024).
\newblock National public data breach: Lawsuit claims failed to protect billions of personal records.
\newblock \url{https://www.itpro.com/security/data-breaches/national-public-data-breach-lawsuit-claims-nearly-three-billion-people-had-personal-data-exposed}.

\bibitem[Kurniawan and Setyawan, 2024]{kurniawan2024importance}
Kurniawan, I.~D. and Setyawan, V.~P. (2024).
\newblock The importance of protecting e-commerce consumer personal data.
\newblock {\em IJOLARES: Indonesian Journal of Law Research}, 2(2):51--55.

\bibitem[Lakshmanan, 2025]{tiktokAliexpress}
Lakshmanan, R. (2025).
\newblock European privacy group sues tiktok and aliexpress for illicit data transfers to china.
\newblock \url{https://thehackernews.com/2025/01/european-privacy-group-sues-tiktok-and.html}.

\bibitem[Langone, 2019]{shoppingApps}
Langone, A. (2019).
\newblock At least 80\% of shopping apps leak users' data. here's how to protect yourself.
\newblock \url{https://money.com/at-least-80-of-shopping-apps-leak-users-data-heres-how-to-protect-yourself/}.

\bibitem[Martiskova and Svec, 2020]{martiskova2020digital}
Martiskova, P. and Svec, R. (2020).
\newblock Digital era and consumer behavior on the internet.
\newblock In {\em Digital Age: Chances, Challenges and Future 7}, pages 92--100. Springer.

\bibitem[Mathur et~al., 2019]{10.1145/3359183}
Mathur, A., Acar, G., Friedman, M.~J., Lucherini, E., Mayer, J., Chetty, M., and Narayanan, A. (2019).
\newblock Dark patterns at scale: Findings from a crawl of 11k shopping websites.
\newblock {\em Proc. ACM Hum.-Comput. Interact.}, 3(CSCW).

\bibitem[Microsoft, 2020]{playwright}
Microsoft (2020).
\newblock Playwright.
\newblock \url{https://playwright.dev/}.

\bibitem[Mori{\'c} et~al., 2024]{moric2024protection}
Mori{\'c}, Z., Dakic, V., Djekic, D., and Regvart, D. (2024).
\newblock Protection of personal data in the context of e-commerce.
\newblock {\em Journal of cybersecurity and privacy}, 4(3):731--761.

\bibitem[Okeke et~al., 2013]{okeke2013issues}
Okeke, R.~I., Shah, M.~H., and Ahmed, R. (2013).
\newblock Issues of privacy and trust in e-commerce: Exploring customers’ perspective.
\newblock {\em Journal of Basic and Applied Scientific Research}, 3(3):571--577.

\bibitem[Pabian et~al., 2020]{pabian2020customer}
Pabian, A., Pabian, B., and Reformat, B. (2020).
\newblock E-customer security as a social value in the sphere of sustainability.
\newblock {\em Sustainability}, 12(24):10590.

\bibitem[Pagey et~al., 2023]{pagey2023all}
Pagey, R., Mannan, M., and Youssef, A. (2023).
\newblock All your shops are belong to us: Security weaknesses in e-commerce platforms.
\newblock In {\em Proceedings of the ACM Web Conference 2023}, page 2144–2154, New York, NY, USA. Association for Computing Machinery.

\bibitem[Papadogiannakis et~al., 2021]{10.1145/3442381.3450056}
Papadogiannakis, E., Papadopoulos, P., Kourtellis, N., and Markatos, E.~P. (2021).
\newblock User tracking in the post-cookie era: How websites bypass gdpr consent to track users.
\newblock In {\em Proceedings of the Web Conference 2021}, WWW '21, page 2130–2141, New York, NY, USA. Association for Computing Machinery.

\bibitem[Rauti et~al., 2024]{rauti2024analyzing}
Rauti, S., Carlsson, R., Mickelsson, S., M{\"a}kil{\"a}, T., Heino, T., Pirjatanniemi, E., and Lepp{\"a}nen, V. (2024).
\newblock Analyzing third-party data leaks on online pharmacy websites.
\newblock {\em Health and Technology}, 14(2):375--392.

\bibitem[Rivers and Lewis, 2014]{rivers2014ethicalresearchstandards}
Rivers, C.~M. and Lewis, B.~L. (2014).
\newblock Ethical research standards in a world of big.
\newblock {\em F1000Research}, 3.

\bibitem[Sakalauskas and Kriksciuniene, 2024]{sakalauskas2024personalized}
Sakalauskas, V. and Kriksciuniene, D. (2024).
\newblock Personalized advertising in e-commerce: Using clickstream data to target high-value customers.
\newblock {\em Algorithms}, 17(1):27.

\bibitem[Similarweb-LTD, 2025]{similarweb}
Similarweb-LTD (2025).
\newblock Similarweb digital intelligence.
\newblock \url{https://www.similarweb.com/}.

\bibitem[Smith, 2024]{temuSued}
Smith, B. (2024).
\newblock Arkansas sues chinese online retailer temu, claims site illegally accessing user information.
\newblock \url{https://www.kark.com/news/state-news/arkansas-sues-chinese-online-retailer-temu-claims-site-illegally-accessing-user-information/}.

\bibitem[Vlachogiannakis, 2025]{openSource}
Vlachogiannakis, I. (2025).
\newblock Open-source data.
\newblock \url{https://github.com/gvlachogiannakis/e-shop-privacy-leaks}.

\bibitem[Weigand, 2023]{backupLeaks}
Weigand, S. (2023).
\newblock Over 12% of online stores accidentally leak data during private backups.
\newblock \url{https://www.scworld.com/news/over-12-of-online-stores-accidentally-leak-data-during-private-backups}.

\end{thebibliography}

\end{document}